# Nonlinear Inverse Synthesis for Optical Links with Distributed Raman Amplification

Son Thai Le, Jaroslaw E. Prilepsky, Paweł Rosa, Juan D. Ania-Castañón and Sergei K. Turitsyn

***Abstract* — Nonlinear Fourier transform (NFT) and eigenvalue communication with the use of nonlinear signal spectrum (both discrete and continuous), have been recently discussed as a promising transmission method to combat fiber nonlinearity impairments. However, because the NFT-based transmission method employs the integrability property of the lossless nonlinear Schrödinger equation (NLSE), the original approach can only be applied directly to optical links with ideal distributed Raman amplification. In this paper, we investigate in details the impact of a non-ideal Raman gain profile on the performance of the nonlinear inverse synthesis (NIS) scheme, in which the transmitted information is encoded directly onto the continuous part of the nonlinear signal spectrum. We propose the lossless path-averaged (LPA) model for fiber links with non-ideal Raman gain profile by taking into account the average effect of the Raman gain. We show that the NIS scheme employing the LPA model can offer a performance gain of 3 dB regardless of the Raman gain profiles.**

## I. Introduction

Continuing demand from the growing number of bandwidth-hungry applications and on-line services (such as cloud computing, HD video streams, on-line content sharing and many others) is pushing the required optical communication system capacity close to the theoretical limit of a standard single-mode fiber (SSMF) [1], which is imposed by the inherent fiber nonlinearity (Kerr effect) [2]. In the last decade, there have been extensive efforts in attempting to surpass the Kerr nonlinearity limit through various nonlinearity compensation techniques, including digital back-propagation (DBP) [3], digital [4] and optical [5-7] phase conjugations (OPCs) at the mid-link or installed at the transmitter [8], and phase-conjugated twin waves [9-11]. However, there are still many limitations and challenges to overcome in applying the aforementioned nonlinear compensation methods, because the transmission technologies utilized in optical fiber communication systems were originally developed for linear (radio or open space) communication channels.

In recent years, there has been a growing interest in the alternative approach of designing fiber optical communication systems in which the nonlinearity is taken into account as an essential element rather than a destructive effect [12]. The core idea behind this approach is based on the fact that without perturbation the nonlinear Schrödinger equation (NLSE), which governs the propagation of optical signal in SSMF, belongs to the class of the so-called integrable nonlinear systems [13-15]. As a consequence, the field evolution over the NLSE channel can be presented within a special basis of nonlinear normal modes (nonlinear signal spectrum), including non-dispersive soliton (discrete) and quasi-linear dispersive radiation (continuous) modes. The evolution of such special nonlinear modes in the fiber channel is essentially linear, which means that the nonlinearity-induced cross-talk between these modes is effectively absent during the propagation (without signal corruption due to noise). Therefore, the parameters of nonlinear modes can be effectively used to encode and transmit information in fiber channel without suffering from nonlinear crosstalk [15-19]. This general idea was first introduced by Hasegawa and Nyu in [12] and was termed there as “eigenvalue communication”.

There are two main directions in the NFT communications methodology, which can be categorized according to what part of the nonlinear spectrum (solitonic discrete part or continuous part) is used for the modulation and transmission. In particular, [19-23] studied the discrete (solitonic) components of the nonlinear spectrum for data communications. This approach is often referred as nonlinear frequency division multiplexing (NFDM) and some initial experimental demonstrations have been reported recently [21-23]. However, the NFDM method requires considerable optimization of the pulse shapes for the purpose of maximizing the resulting spectral efficiency (SE) [24]. The second approach based on the modulation of the continuous part of the nonlinear spectrum, has been proposed in [17] and assessed in detail in [25, 26] (for ideal Raman amplification and EDFA-based optical links) – and was termed there as the nonlinear inverse synthesis (NIS) method. Finally we note that there already exists a very recent study where both the continuous and discrete parts of the nonlinear spectrum have been used simultaneously [27]. However, all transmission schemes employing the NFTs are based on the integrability of the lossless NLSE and thus can be effectively applied only to optical links with ultra-long fiber laser-based distributed Raman amplification providing a flat quasi-lossless gain profile [28].

Research supported by Engineering and Physical Sciences Research Council (EPSRC) through the project UNLOC (EP/J017582/1). This work was presented in part at the 2015 European Conference on Optical Communication (ECOC), Sep. Valencia, Spain.

S. T. Le, J. E. Prilepsky and S. K. Turitsyn are with Aston Institute of Photonic Technologies (AIPT), Aston Triangle, Birmingham, B4 7ET, UK (corresponding author phone: +44(0)744-702-40-09; e-mail: let1@aston.ac.uk). P. Rosa and J. Ania-Castañón are with Instituto de Optica, IO-CSIC, CSIC, Madrid, 28006, Spain.

In this paper, we extend the results firstly presented in [29] and discuss in details the impact of the non-ideal Raman gain profile on the performance of NIS-based transmission systems. We introduce a LPA NIS scheme which offers 3 dB performance gain regardless of the particular Raman profile. To demonstrate the effectiveness of the LPA NIS scheme, without loss of generality, we consider here open-cavity random distributed feedback (DFB) laser Raman amplification, as this scheme provides the best performance among various other Raman amplification schemes [30].

The remainder of the paper is organized as follows. The basic of NFT-based transmission method including system designs is reviewed in Section II. In Section III the design and characteristics of Random DFB laser Raman amplifier is presented. In Sections IV, the concept of LPA model and modified NIS scheme accounting for non-ideal gain profile are introduced. The simulation setup, results and discussion is presented in Section V. Section VI concludes the paper.

## II. Basics of NFT-based Transmission Method

As mentioned earlier, NFT-based transmission employs the nonlinear signal spectrum (discrete and/or continuous parts) for the purpose of data modulation and transmission in fiber optical communication links. This approach can be realized in various ways. Herein we recall the basics of NFTs and different concepts of the NFT-based systems.

### A. Basics of NFT operation

According to the inverse scattering transform (IST) theory, the propagation of a complex signal $q(z, t)$ in nonlinear integrable systems, such as the one governed by the lossless normalized NLSE (Eq. 1), can be decomposed into the linear propagation of its nonlinear spectral data (nonlinear spectrum) [14]. We write the normalized NLSE (the anomalous dispersion case) as

$$jq_z + \frac{1}{2} q_{tt} + q|q|^2 = 0, \qquad (1)$$

where $z$ stands for the propagation distance and $t$ is the time in the frame co-moving with the group velocity of the envelope.

According to the IST theory, the solution of lossless NLSE (1) with a given input signal $q(0, t)$ can be found through three basic steps: i) Obtaining the nonlinear spectral data of the input signal through the NFT. ii) Propagating the nonlinear spectral data to a desired distance, and this propagation is trivial and linear. iii) Defining the output signal through the inverse NFT (INFT) given the evolved nonlinear spectral data at the desired distance $z$. Basically, the NFT converts the signal into the correspondent nonlinear spectrum, including both continuous and discrete parts. This operation is achieved by solving the spectral Zakharov–Shabat problem (ZSP) [14], which corresponds to a scattering problem for a non-Hermitian Dirac-type system of equations for two auxiliary functions $v_{1,2}(t)$, with the NLSE input pulse profile, $q(0,t) \equiv q(t)$, serving as an effective potential:

$$\frac{dv_1}{dt} = q(t)v_2 - j\varsigma v_1, \quad \frac{dv_2}{dt} = -\bar{q}(t)v_1 + j\varsigma v_2 \qquad (2)$$

Here, $\zeta$ is a (generally complex) eigenvalue, $\zeta = \xi + j\eta$, $\bar{q}(t)$ is the complex conjugation of the potential $q(t)$, which is assumed to decay as $t \to \pm\infty$ (technically, we simply truncate our $q(t)$ to a finite duration).

To define nonlinear spectral data, for real $\zeta = \xi$ we can select two specific linearly-independent solutions of Eq. (2) as

$$\Phi(t,\xi) = [\phi_1, \phi_2]^T, \quad \Phi^*(t,\xi) = [\bar{\phi}_2, -\bar{\phi}_1]^T,$$

with the condition at the trailing end of our input profile

$$\Phi|_{t\to-\infty} = [e^{-j\xi t}, 0]^T$$

Then, the two Jost scattering coefficients (NFT spectral amplitudes) $a(\xi)$ and $b(\xi)$ are given by:

$$a(\xi) = \phi_1(t,\xi)e^{j\xi t}|_{t\to\infty}, \; b(\xi) = \phi_2(t,\xi)e^{-j\xi t}|_{t\to\infty}$$

After defining the Jost scattering coefficients, the continuous part of the signal nonlinear spectrum (the “left” reflection coefficient) is defined as:

$$r(\xi) = \bar{b}(\xi) / a(\xi), \qquad (3)$$

The solitons (discrete part) correspond to the complex eigenvalues $\zeta_n = \xi_n + j\eta_n$ in the upper-half of the complex plane, where $a(\zeta_n) = 0$. Together with the eigenvalue $\zeta_n$, each solitonic degree of freedom is characterized by the second complex quantity, the so-called norming constant:

$$C_n = \bar{b}(\varsigma_n) / a'(\varsigma_n), \qquad (4)$$

In general, the NFT maps the initial field, $q(0,t)$, onto a set of scattering data $\Sigma = [(r(\xi), \xi \text{ is real}); (\zeta_n, C_n)]$, where the index $n$ runs over all discrete eigenvalues of the ZSP (if the latter are present) (Fig. 1(a)). More detailed explanations on NFT can be found in [19, 25]. Numerical methods for NFT can be found in [25, 31, 32] and the references therein.

### B. Inverse NFT operation

The inverse INFT maps the scattering data Σ onto the field $q(t)$, Fig. 1(b): This is achieved via the solution of Gelfand–Levitan Marchenko equations (GLME) for the unknown functions $A_{1,2}(x, t)$ [14, 17, 25]:

$$\begin{aligned} &A_1(x,t) + \int_{-\infty}^{x} F(t+y)\bar{A}_2(x,y)dy = 0 \\ &A_2(x,t) - \int_{-\infty}^{x} F(t+y)\bar{A}_1(x,y)dy = F(x+t), \quad x > t \end{aligned} \qquad (5)$$

The quantity F($t$) can contain contributions from the solitonic (discrete) and radiation (continuous) spectrum parts:

$$F(t) = F_{slo}(t) + F_{rad}(t), \qquad (6)$$

$$F_{slo}(t) = -j\sum_n C_n e^{-j\varsigma_n t}, \qquad F_{rad} = \frac{1}{2\pi}\int_{-\infty}^{+\infty} r(\xi)e^{-j\xi t}d\xi$$

Having solved the GLME (5), the output of the INFT, i.e. the field profile in the true space-time domain, is given by

$$q(t) = 2\lim_{x\to t-0} A_2(t,x) \qquad (7)$$

More detailed explanations on INFT can be found in [19, 25]. Effective numerical methods for INFT (with discrete and/or continuous parts) can be found in [25, 32, 33] and references therein.

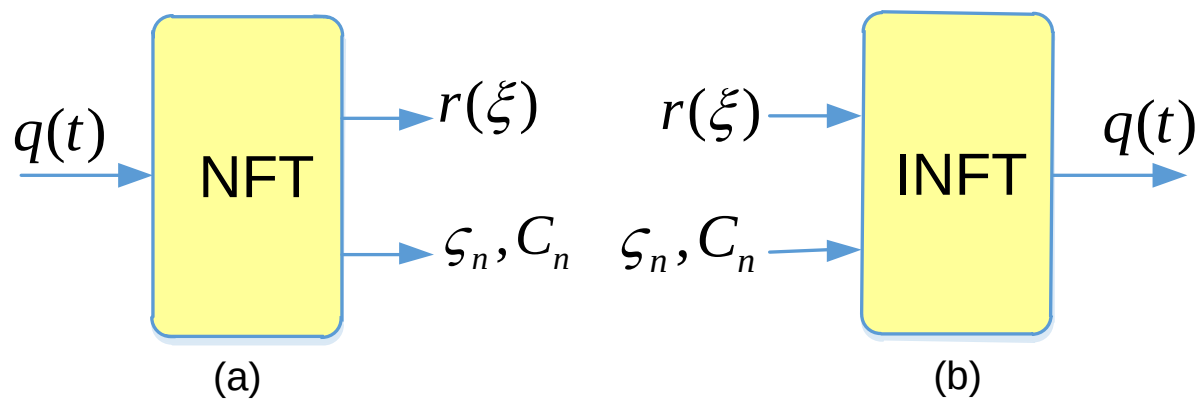

Fig. 1. Illustration of NFT and INFT for a given input the potential $q(t)$, which is assumed to decay as $t \to \pm\infty$

### C. Evolution of the nonlinear spectrum

When one is interested in the solution $q(L, t)$ at the distance $z = L$, the quantities $r(\xi)$, $C_n$, $\zeta_n$ in Eq. (3-4) are replaced with $r(z, \xi)$, $C_n(z)$ $\zeta_n(z)$. Under the noise-free assumption, the evolution of nonlinear spectrum in lossless NLSE channel is trivial and linear:

$$r(z,\xi) = r(0,\xi)\cdot e^{-2j\xi^2 z},\ \varsigma_n(z) = \varsigma_n,\ C_n(z) = C_n(0)\cdot e^{-2j\varsigma_n{}^2 z} \quad (8)$$

So, each discrete eigenvalue $\zeta_n$ is an integral of motion and does not change, while the reflection coefficient and norming constant obey a simple dynamics, similarly to the evolution of ordinary Fourier modes in a linear dispersive channel. This remarkable property makes nonlinear spectrum (discrete and/or continuous parts) ideal information carriers in nonlinear fiber channels.

### D. NFT-based system designs

As the evolution of nonlinear spectrum is essentially linear in nonlinear lossless fiber channel, the nonlinear spectrum can be used for data modulation and transmission. The two basic designs for NFT-based transmission systems are presented in the Fig. 2.

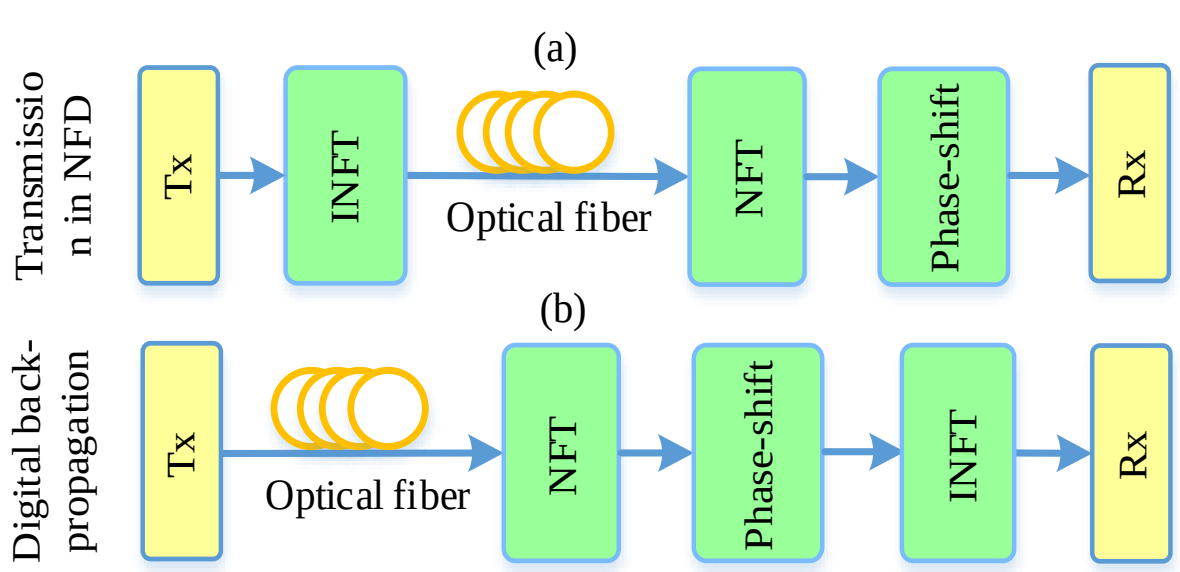

Fig. 2. Basic designs of NFT-based transmission systems, including transmission in the nonlinear Fourier domain (NFD) (a) and digital back propagation in NFD (b).

The first design (Fig. 2(a)) can be referred as transmission in the nonlinear Fourier domain (NFD) as the transmitted information is encoded directly onto the nonlinear signal spectrum (discrete and/or continuous parts) via the INFT. Herein, NFT provides three degrees of freedom for data modulation, namely the discrete spectrum ($C_n$), continuous spectrum ($r(\xi)$) and discrete eigenvalues ($\zeta_n$). The modulations of discrete spectrum, continuous spectrum and discrete eigenvalues are often considered separately due to the numerical complexity of the full NFT-INFT cycle. The resulted transmission methods can be termed as NFDM, NIS and eigenvalue communication, respectively. Herein, we focus on the NIS scheme as it can be combined effectively with high modulation format and traditional coherent transmission technologies, offering highest flexibility in the system design [25, 26]. In the second design, the NFTs are used to cancel the nonlinearity distortion in fiber optical communication systems. This scheme can be understood as the digital back propagation with the use of the NFT operations, NFT-DBP, Fig. 2(b) [34]. Here, the signal encoding and modulation is performed in the space-time domain. As the evolution of the nonlinear spectrum is linear, the interplay of dispersion and nonlinearity during transmission can be removed using a single phase-shift operation. After removing the linear phase shift inside the NFD, the transmitted signal can be recovered via the INFT.

In general, both aforementioned system designs can be effectively applied to cancel the deterministic nonlinear distortions in fiber optical links. However, from a practical viewpoint, the first design provides several advantages over the second NFT-DBP design. First of all, it admits the flexibility of choosing what part of nonlinear spectrum one employs for the data modulation, and thus, significantly reduces the numerical complexity associated with the full INFT through appropriate modulation techniques. Second, in the first design the computational load can be effectively split between the transmitter and receiver sides. On the other hand, the second design requires modification only at the receiver side, while modifications are necessary at both sides in the first design.

## III. Random DFB Raman Amplifier

As discussed above, the NFT-based methods can be applied directly only to lossless or quasi-lossless optical links. In practice, such condition is difficult to achieve. As a result, it is important to investigate the impact of non-flat gain profile on the performance of NFT-based systems and develop appropriate modification of the NFT approach. Herein, we take into account an open-cavity random DFB laser Raman amplification scheme, which can provide various gain profiles by controlling the forward pump power (FPP) [30].

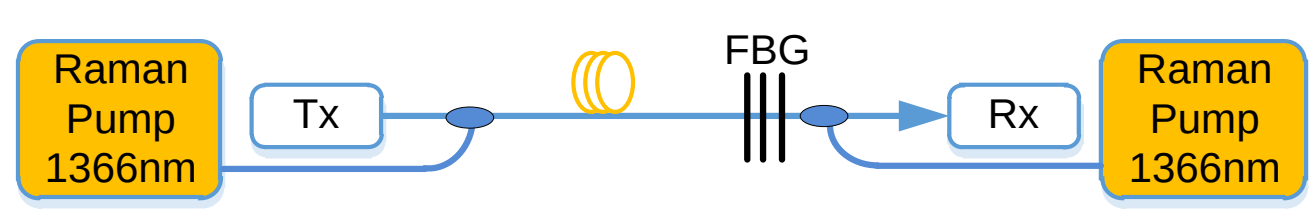

Fig. 3. Schematic of the random DFB laser Raman amplifier

The schematic design of the random DFB laser Raman amplifier that allows achieving 2nd order pumping with a single wavelength pump is shown in Fig. 3. In this scheme, a high reflectivity (99%) fiber Bragg grating (FBG) centered at 1455 nm with 200 GHz bandwidth is deployed at the end of the transmission span to reflect back-scattered Rayleigh Stokes-shifted light from the backward pump (at 1366 nm) and stimulate random DFB lasing at 1455nm (wavelength of the FBG). This random DFB laser acts as the first order pump, together with 2nd order pump, to amplify the signal at 1550 nm. The lack of an FBG on the side of the forward pump significantly reduces the relative-intensity-noise transfer from the forward pump to the Stokes-shifted light at 1455 nm [35], which can seriously hinder coherent transmission [36].

We simulated the signal and noise power excursion for different pump power ratios in Raman amplifiers using the experimentally verified model [28] with an appropriate boundary conditions that shows a high degree match with the OTDR traces. In all cases the considered Raman pumps are fully depolarized. The backward pump powers were chosen accordingly to provide a net gain of 0 dB. The simulated gain and noise profiles along 80 km length SMF span are shown in Fig. 4 for different FPPs. It can be seen in Fig. 4 that, when the FPP is increased, the Raman gain increases while the noise

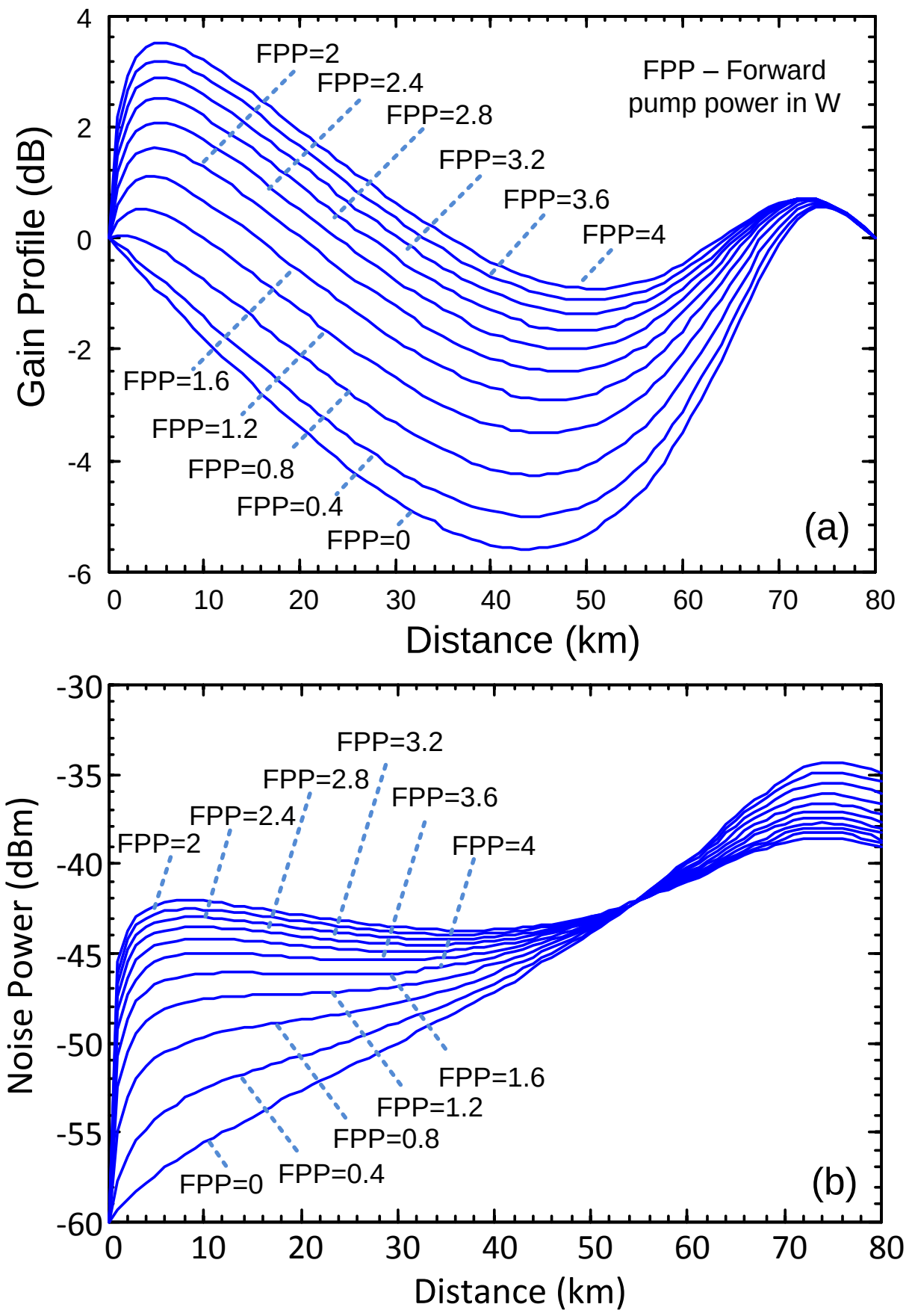

Fig. 4. Raman gain (a) and noise (b) profiles along 80km SMF span for different value of the forward pump power (FPP).

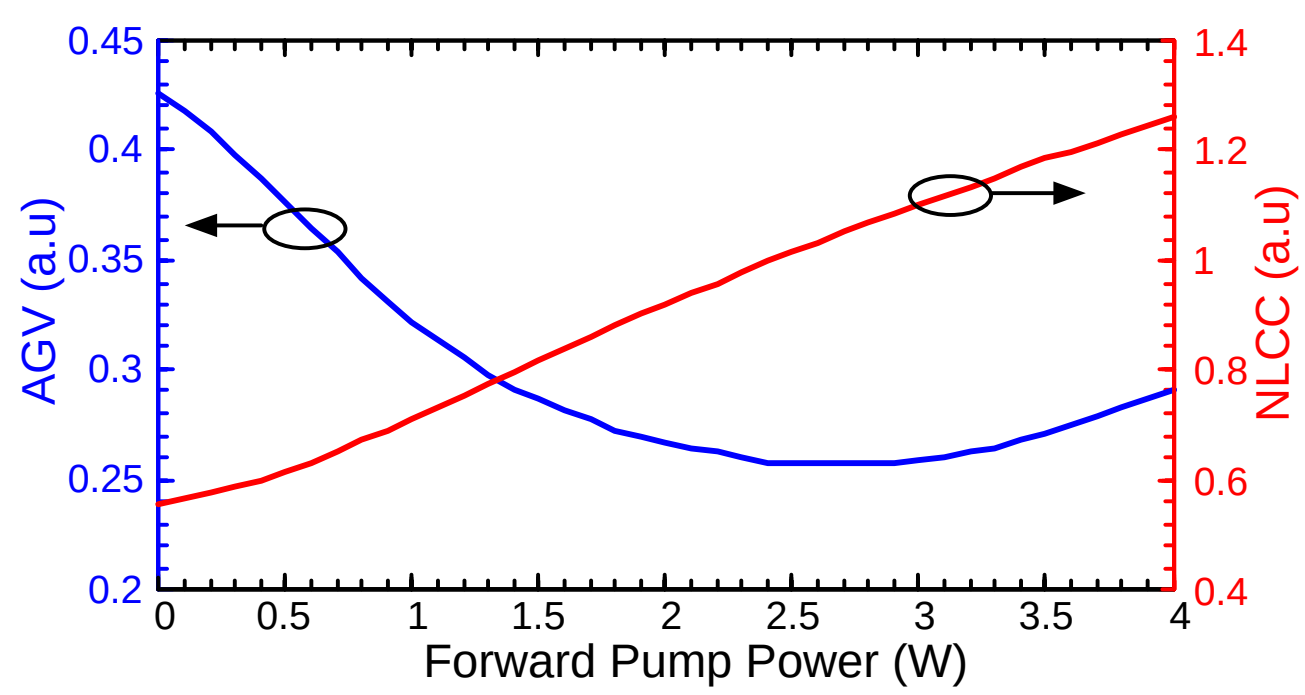

Fig. 5. AGV and NLCC as a function of the forward pump power; the span length was 80km.

power decreases, leading to a better signal-to-noise ratio.

To characterize the non-flatness of the Raman gain profile, we define the average gain variation (AGV) of the Raman gain profile as:

$$AGV = \int_{z=0}^{L} \left| G(z) - k \right| dz / \left( L G_0 \right), k = \int_{z=0}^{L} G(z) dz / L, \qquad (9)$$

where G($z$) is the Raman gain normalized to 1 at the beginning of the span, $L$ is the span length, and $k$ is termed as the nonlinear correction coefficient (NLCC).

The AGV is shown in Fig. 5 as a function of the FPP. The FPP = 0 case corresponds to the backward-pumping-only scheme. From Fig. 5 we see that the AGV can be effectively reduced by increasing the FPP (up to its optimum value) at the cost of increasing the system power consumption. The optimum value of FPP was found to be ~2.7 W, giving the AGV ≈ 0.255. By varying the FPP to vary the AGV we can effectively investigate the impact of Raman gain flatness on the NIS-based transmission systems.

## IV. Modified NIS for Non-ideal Distributed Raman Amplification

As discussed in the previous section, for practical Raman amplification schemes, the non-flatness level characterized by the AGV, can be as high as ~ 0.43 (for backward-pumping-only scheme). This high level of non-flatness may deprive all the nonlinearity cancellation benefit of NIS and of other NFT-based transmission schemes. As a result, the LPA model for Raman-based optical links should be developed in a similar manner to EDFA-based optical links [26] in order to apply NIS and, potentially, other NFT-based transmission schemes.

The general model of the NLSE for optical links with Raman amplifiers can be written as

$$jq_z - \beta_2 q/2 + \gamma q|q|^2 = jg(z)q, \qquad (10)$$

where $t$ is the time in the frame co-moving with the group velocity of the envelope, $\beta_2 < 0$ is the dispersion coefficient, $\gamma$ is the Kerr nonlinearity coefficient and $g(z)$ is the distributed distance-dependent Raman gain coefficient. We consider here the case that the same pumping scheme is applied to all fiber spans. In this case $g(z)$ is a periodic function with a period equal to the span length. By introducing the standard change of variables [37]:

$$q(z,t) = \exp\left( \int_0^z g(y) dy \right) A(z,t),$$

Eq. (10) can be rewritten as:

$$jA_z - \beta_2 A/2 + \gamma G(z) A|A|^2 = 0, \qquad (11)$$

which is the lossless NLSE with a distance-dependent nonlinear coefficient; here G(z) is the instantaneous gain

$$G(z) = \exp\left( 2\int_0^z g(y) dy \right)$$

We assume here that the dynamic of the envelope $A(z, t)$ does not change significantly after each fiber span. In this case, the distance-dependent nonlinear coefficient in (11) can be replaced by its averaged value over each fiber span, giving the effective LPA NLSE [37]

$$jA_z - \beta_2 A/2 + \gamma \left( \int_0^L G(z) dz / L \right) A|A|^2 = 0 \qquad (12)$$

From an engineering point of view, the LPA NLSE model can be obtained from the general NLSE model by removing the loss term and updating the nonlinear coefficient in such way that the nonlinear phase-shift acquired by the signal during propagation over one span is unchanged. Of course, as shown in [26] the LPA NLSE is an approximated model so its accuracy depends strongly on the signal and system parameters such as bandwidth, pulse shape, power and transmission distance. A detailed investigation on the accuracy of LPA NLSE model for optical links with lumped amplifications can be found in [26]. For the Raman amplification, simulation results (not shown here) indicate very similar dependence of the accuracy of LPA model on signal's and system's parameters in comparison to the case of lumped amplification.

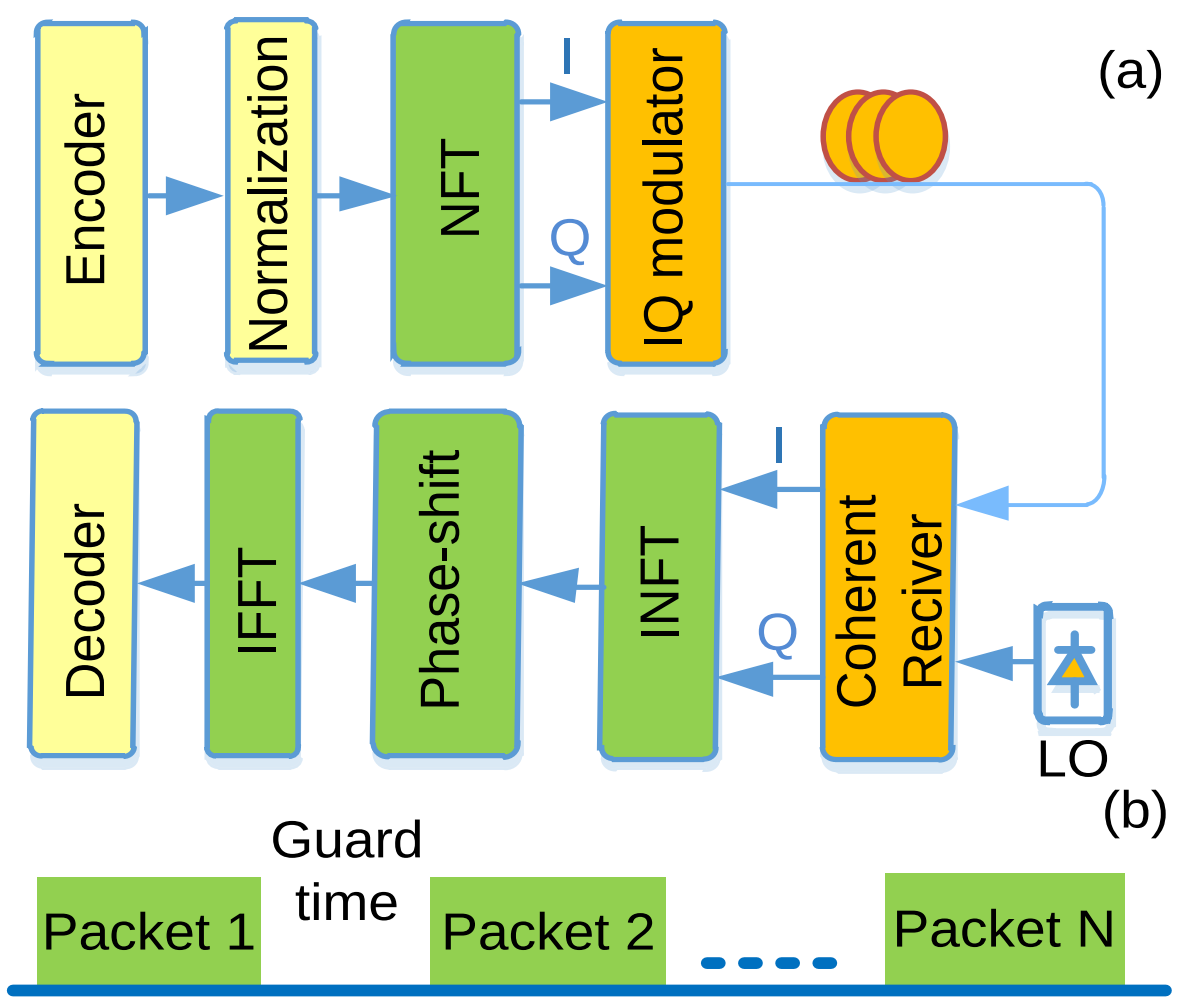

Fig. 6(a) – Block diagram of NIS-based transmission systems, (b) – Illustration of a burst mode transmission

Based on the obtained LPA NLSE model, we can design the appropriate modified NIS scheme accounting for the non-ideal Raman gain profile as shown in Fig. 6(a). Firstly, the normalization is performed on the initial signal $q(t)$ (modulated with arbitrary pulse shape and modulation format) to bring the LPA NLSE model to the standard normalized form (Eq. 1)

$$t/T_S \to t, z/Z_S \to z, q\sqrt{\gamma \cdot k \cdot Z_S} \to q, \quad (13)$$

where the time normalization $T_s$ is a free parameter (e.g., a characteristic time scale or a reciprocal bandwidth) and the associated space scale is $Z_s = T_s^2/|\beta_2|$; $k$ is NLCC defined as in Eq. (9). The dependence of the NLCC on FPP is shown in Fig. 5 (red curve).

After the normalization, the linear Fourier spectrum of the encoded input waveform ($Q(\omega)$) is mapped onto the continuous part of the nonlinear spectrum of another signal ($s(t)$) to be transmitted using the INFT. Let $r(\xi)$ denote the continuous part of the nonlinear spectrum of $s(t)$ then the mapping operation of the INFT block can be expressed as:

$$r(\xi)\big|_{\xi=-\omega/2} = -Q(\omega) \quad (14)$$

This defines the pre-processing of the initial signal at the transmitter and can be considered as nonlinearity pre-compensation. Generally, the mapping rule between $Q(\omega)$ and the reflection coefficient $r(\xi)$, Eq. (3), can be taken arbitrarily. However, the usage of mapping given by Eq. (14), utilized in the series of previous works on the NIS method [17, 25-26], ensures the exact convergence of the nonlinear quantities to the respective linear analogs in the low power limit, which simplifies the system design and checks.

The INFT-generated complex signal, $s(t)$, is then fed into the IQ modulator for direct up-converting into the optical domain and launched into the fiber. At the receiver, the real and imaginary parts of the transmitted signal are detected with a coherent receiver. The nonlinear spectrum of the received signal is then obtained by using the NFT. As the evolution of the signal nonlinear spectrum is linear and trivial within the LPA NLSE model, Eq. (8), the linear Fourier spectrum of the initial encoded complex signal can be recovered by applying a single step linear phase-shift removal

$$\overline{Q}(\omega) = -r(L,\xi) \cdot e^{2j\xi^2 L}\big|_{\xi=-\omega/2} \quad (15)$$

Then, having unrolled the dispersion-induced phase shift, the initial encoded waveform $q(t)$ can be recovered using the IFFT operation and, finally, it can be fed into the standard decoder for data detection. In general, the DSP at the receiver of an NIS-based system involves a single NFT operation and a single linear compensation step to remove the nonlinear impairments without reverse propagation, independently of the transmission distance. This is a significant advantage of the NIS method over the other nonlinear compensation techniques.

## V. Simulation Setup and Results

As discussed in [25, 26], the NIS transmission method can be combined with any modulation formats and transmission schemes. A comparison of OFDM and single carrier with Nyquist pulse shaping (or orthogonal time division multiplexing) for NIS-based systems was provided in [25], revealing that the OFDM is a more suitable modulation format because it provides a smaller $L1$-norm, thus diminishing the processing signal cut-off error. This can be explained by the fact that the L1 norm of the Fourier transform is always lower or equal to the $L1$-norm of the time-domain signal. As a result, in this paper we consider only the OFDM scheme.

We design here the 16QAM 56-Gbaud OFDM NIS-based systems in the burst mode regime (Fig. 6(b)), as the NFT operations have to be performed on return-to-zero signals. In this scheme, the neighbouring packets are separated by a guard time, which is 20% longer than the dispersion induced memory. The dispersion induced memory of the link can be estimated as:

$$\Delta T = 2\pi B \beta_2 L, \quad (16)$$

where $B$ is the signal's bandwidth and $L$ is the link distance.

For simplicity, we assume that each packet data contains only one OFDM symbol. To generate the OFDM signals, the IFFT size of 1024 was used, where 112 subcarriers were filled with data (with Gray-coding) while the remaining subcarriers were set to zero for oversampling purpose. The useful OFDM symbol duration is 2ns. No cyclic prefix was added to the signal. The net data rate, after removing 7% overhead due to the FEC, was 200 Gb/s (considering only the burst's bit-rate). Herein, we aim to show that DSP techniques based on model (12) can be applied effectively even in the long-haul optical communication systems with as large a bandwidth as 56 GHz.

The propagation of signal in fiber link was simulated using the split-step Fourier method with a step size of 1 km, using the gain profile shown in Fig. 4(a). The Raman noise was modelled as a Gaussian noise, which was added to the signal after each step (1 km), following the simulated noise profiles shown in Fig. 4(b). The system performance was evaluated through the EVM and the estimated BER was then converted to the Q-factor [38].

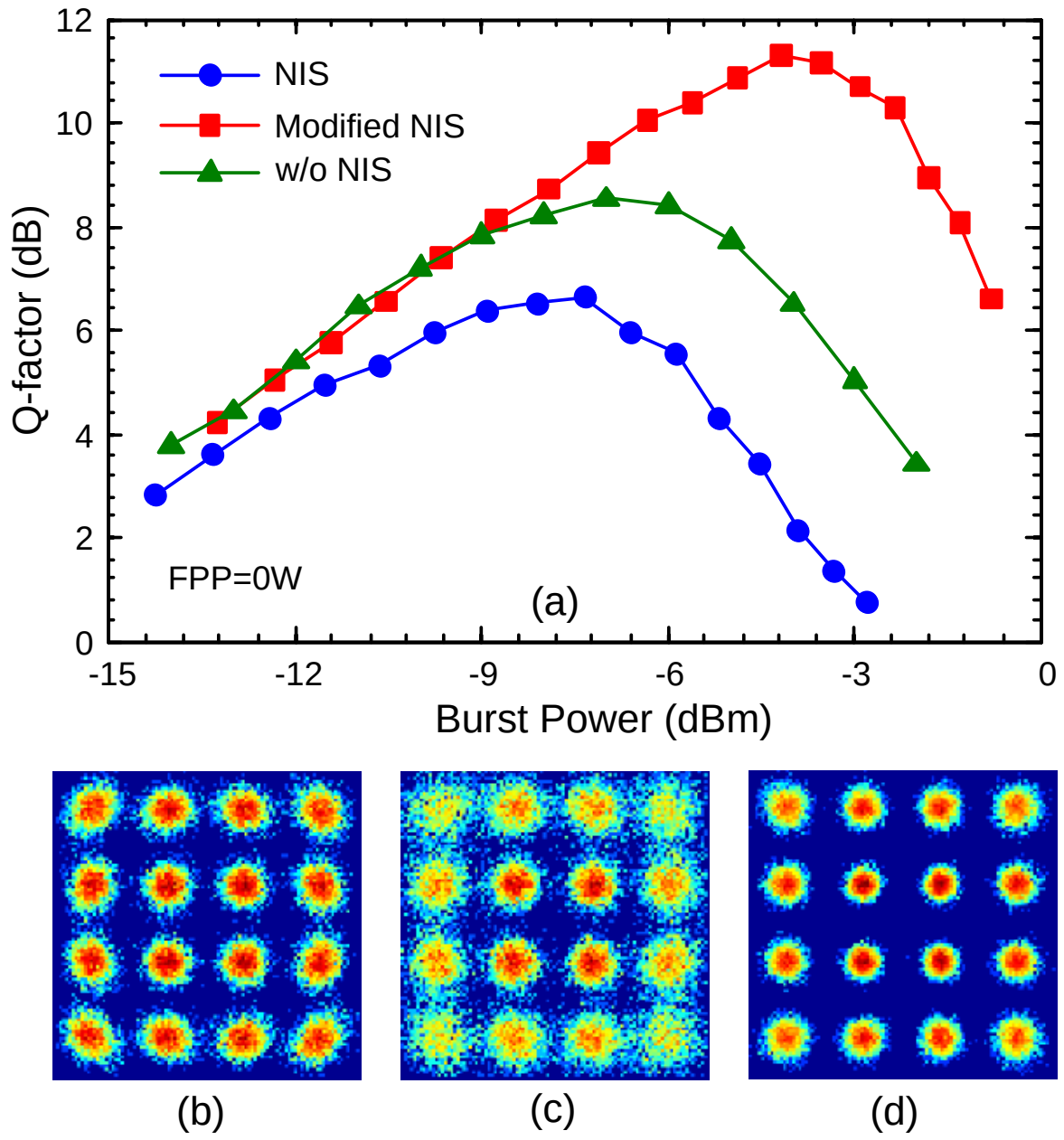

Fig. 7(a). Performance comparison of OFDM systems with and without NIS, and the modified NIS scheme for FPP =0 W (backward pumping only case), (b)-(d) constellations at the optimum launch powers for system without NIS (b), with NIS (c) and with the modified NIS (d) schemes at FPP=0W, respectively.

The performance of OFDM systems with and without NIS, and with the proposed modified NIS scheme is compared in Fig. 7(a) for the backward pumping only scheme (FPP=0 W). For the case of backward pumping scheme (FPP = 0 W), because of the high non-flatness level of the Raman gain profile (AGV ~ 0.43), applying directly the NIS method worsens the system performance by ~ 2 dB. This result clearly indicates that the non-ideal Raman gain profile has a significant impact on the NIS-based systems: When the AGV is high (AGV~0.43 if FPP=0W), the NIS method cannot produce any advantage due to the wrong power estimate. However, if the modified NIS scheme is employed, a Q-factor improvement of ~3 dB is observed. This effectively means that the performance of the NIS scheme is enhanced by 5 dB by simply employing the NLCC that takes into account the non-ideal gain profile along the span. The received constellations at optimum launch powers for three systems under investigation are shown in Figs. 7(b)-(d).

A similar performance comparison result is plotted in the Fig. 8 when FPP =2.7 W, which provides the smallest level of non-flatness of the Raman gain profile. It should be noted here that by increasing the FPP, the Raman noise figure is reduced, which can be referred from the performance at the low power level. It can be seen in Fig. 8 that the NIS method gives around 2 dB performance gain if FPP = 2.7W. In this case, the AGV is relatively small (0.255), and the NIS scheme still offers a meaningful performance gain. However, if one uses the modified NIS method based on Eq. (12), an extra ~1 dB gain can be achieved, giving a total performance gain of ~3 dB.

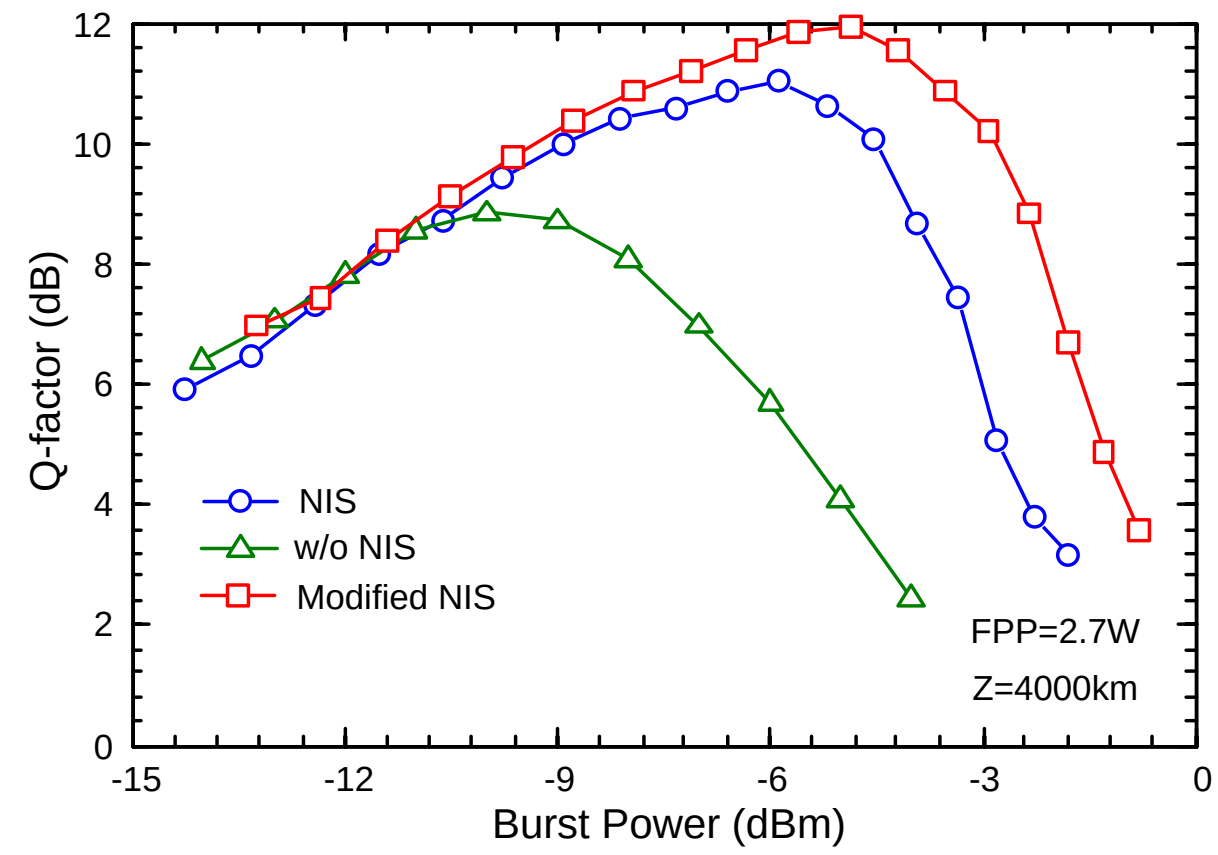

Fig. 8. Performance comparison of OFDM systems with and without NIS, and the modified NIS scheme for FPP =2.7 W. The transmission distance is 4000 km

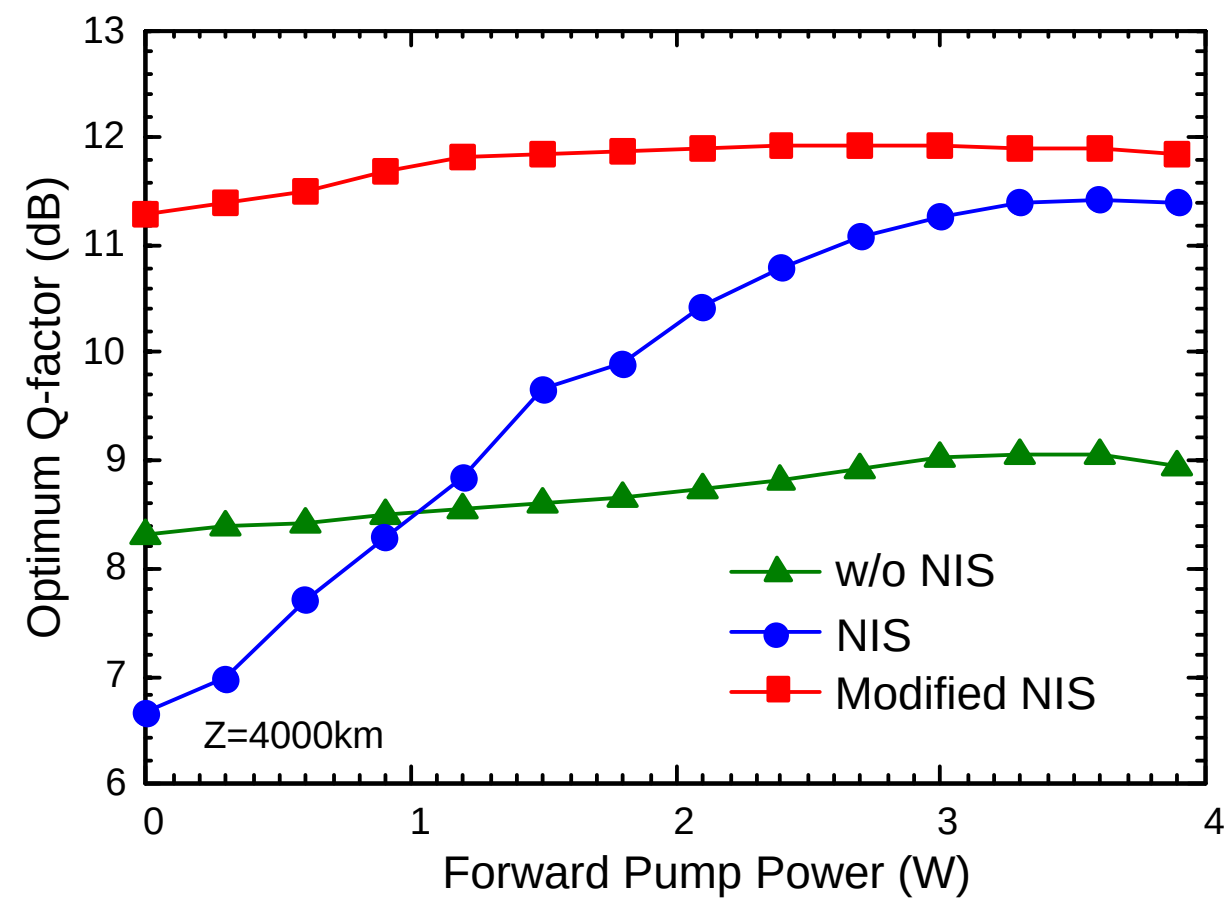

Fig. 9. Optimum Q-factor as a function of FPP for OFDM systems with and without NIS, and the modified NIS scheme. The transmission distance was 4000km

The optimum Q-factors in systems with and without NIS and with the modified NIS schemes are presented in Fig. 9 as functions of the FPP. As expected, when the FPP is increased from 0 W to 2.7 W, the achievable performance of NIS-based system increases dramatically as a result of the decrease in the AGV. However, increasing further the FPP, which increases the AGV accordingly (if FPP > 2.7 W), does not decrease the performance of NIS-based system. We attribute this phenomenon to the reduction of the amplifier noise figure when the FPP is increased. On the other hand, the modified NIS scheme offers around 3 dB gain in Q-factor, independently of the FPP. This indicates that if the modified NIS method is applied, the forward pump is not necessary, which offers a significant reduction in the cost and power consumption when designing the NIS-based systems with Raman amplifications.

## VI. Conclusion

We have shown that the non-ideal Raman gain has a significant impact on the performance of NIS-based systems. As a result, an appropriate modification is required in order to apply NIS techniques in optical links with practical Raman amplification schemes. Based on the LPA NLSE model, which takes into account the average effect of the Raman gain profile, we have proposed a modified NIS scheme for optical links with Raman amplifications. By considering various designs of random DFB laser Raman amplifier we have shown that the modified NIS scheme can offer a performance gain of 3dB, independently of the Raman gain profile. We strongly believe that the LPA NLSE model can be effectively applied in different other NFT-based transmission systems. Moreover, it seems possible to attain the further improvement of the LPA NLSE model performance, which would take into account higher order corrections to Eq. (12), similarly to the higher-order guiding-center models proposed in [13] for the lumped EDFA amplification method.

## ACKNOWLEDGMENT

This work was supported by the UK EPSRC Grant UNLOC (EP/J017582/1). P. Rosa kindly acknowledges the support of the EU Marie Skłodowska-Curie IF CHAOS (658982) project.